\documentclass[twocolumn,aps,prc,superscriptaddress,showpacs,floatfix]{revtex4}
\usepackage{graphicx}

\begin{document}

\newcommand{\beq}{\begin{equation}}
\newcommand{\eeq}{\end{equation}}
\newcommand{\eg}{e.g.}
\newcommand{\etal}{{\it et.~al.}}
\newcommand{\ie}{i.e.}

\title{Quark Coalescence for Charmed Mesons in 
       Ultrarelativistic Heavy-Ion Collisions} 

\author{V.~Greco}
\author{C. M. Ko}
\author{R. Rapp}
\affiliation{Cyclotron Institute and Physics Department, Texas A\&M 
University, College Station, Texas 77843-3366}

\date{\today}

\begin{abstract}
We investigate effects of charm-quark interactions in a 
Quark-Gluon Plasma on the production of $D$ and $J/\psi$ mesons in 
high-energy heavy-ion collisions.  Employing a previously constructed 
coalescence model that successfully reproduces the transverse momentum 
($p_T$) spectra and elliptic flow ($v_2(p_T)$) of light hadrons at 
RHIC from underlying light-quark distributions at the phase transition
temperature $T_c$, $D$-meson and $J/\psi$ $p_T$-spectra are evaluated. 
For the charm-quark distributions, we consider two limiting scenarios: 
(i) {\em no} rescattering, corresponding to perturbative QCD (pQCD) 
spectra and (ii) {\em complete} thermalization including transverse
expansion. 
We find that $D$-mesons acquire a minimal $v_2$ inherited from their 
light-quark content and corresponding semileptonic decay spectra of 
single electrons practically preserve the $v_2$ of the parent particles,
exhibiting marked differences between the pQCD and thermal scenarios 
for $p_T\ge 1$ GeV.
Likewise, the $p_T$-spectra and yields of $J/\psi$'s differ 
appreciably in the two scenarios.
\end{abstract}

\maketitle

\section{Introduction}
Collisions of heavy nuclei at the Relativistic Heavy-Ion Collider 
(RHIC) are expected to provide conditions favorable for the creation 
of a deconfined and chirally symmetric Quark-Gluon Plasma (QGP). 
Among the promising probes of this phase are hadrons
containing charm ($c$) quarks. On the one hand, the abundance of 
observed $J/\psi$ mesons was predicted to be sensitive to QGP 
formation~\cite{MS86} due to Debye screening of the heavy-quark 
potential, leading to suppression of its production.  On the other 
hand, spectra of open-charm states (mostly $D$-mesons) are believed 
to encode valuable information on charm-quark reinteractions in 
the hot and dense medium.  Due to the relatively large charm-quark 
mass, $m_c\simeq 1.5$ GeV, $c\bar c$ production is presumably dominated 
by primordial $N$-$N$ collisions \cite{muller}, whereas thermalization 
of their momentum distributions is still an open issue, with important 
ramifications for charmonium production.  

At RHIC energies, with an expected 10-20 $c\bar c$ pairs per central 
$A$-$A$ collision, statistical models predict a substantial regeneration 
of  $J/\psi$'s by recombination of $c$ and $\bar c$ quarks close to the 
phase transition~\cite{pbm00,thews01,goren01,GR01,thews03,GRB03}. These 
estimates reside on the assumption that charm quarks are in thermal 
equilibrium with the surrounding medium. Recombination approaches 
have also been applied before with fair phenomenological success 
in studying flavor dependencies of open-charm production in elementary 
$p$-$N$ and $\pi$-$N$ collisions~\cite{Hwa95,Bra03,RS03}.
Again, the heavy quark serves as a probe for the (nonthermal) 
hadronization environment via coalescence-type processes with 
surrounding valence or sea quarks. Finally, parton coalescence 
within a hadronizing QGP  has recently been put forward as a 
mechanism for light hadron production at intermediate $p_T\simeq$~2-6~GeV 
in heavy-ion collisions at RHIC. The observed ``anomalous" $\bar p/\pi$ 
ratio of $\sim$~1, as well as the apparent ``constituent-quark" scaling 
of the hadron $v_2$, are naturally accounted for within this 
framework~\cite{greco,greco2,Hwa03,fries,fries2,volo}.

In this paper, we employ quark coalescence to evaluate 
spectra of open- and hidden-charm mesons at RHIC, 
with the objective to address the following issues: 
(a) What is the sensitivity of the $J/\psi$ abundance and $p_T$
spectrum to the underlying momentum distribution of charm quarks? 
(b) How does the interplay of the charm- and light-quark 
distributions translate into the $p_T$-spectrum and elliptic flow of 
$D$-mesons? Point (a) lifts the assumption of complete $c$-quark 
thermalization common to statistical models, which is important to 
discriminate regeneration from suppression mechanisms.    
A first study of this kind has been performed in Ref.~\cite{thews03} 
where $c$-quark distributions from pQCD have been implemented into a 
kinetic rate equation solved in the background of an evolving QGP.
For $c$-quark rapidity densities of $dN_{c\bar c}/dy\sim 2.5$ in central 
Au+Au collisions at RHIC, the final $J/\psi$ number has been found 
to deviate from scenarios with thermalized $c$-quarks by $\sim$~50\%.  
Our results to be discussed below differ from these 
estimates.    
Concerning point (b), the recent study of Ref.~\cite{Bats03} has 
shown that single-electron $p_T$ spectra from decays of PYTHIA-generated 
and hydrodynamic $D$-meson distributions cannot by discriminated 
by current PHENIX data~\cite{phenixe}, extending to $p_T^e\simeq$~3~GeV.
In the present work, $D$-mesons are formed from $c$-quark coalescence
with thermal light quarks at hadronization, where the latter also carry
the collective expansion characteristics (radial and elliptic flow)
that underlies a satisfactory description of pion and baryon
spectra in Refs.~\cite{greco,greco2}. An important point in the
present study will be the evaluation of the elliptic flow of both $D$-mesons
and their electronic decay spectra. For both (a) and (b) we will 
consider two limiting cases for the $c$-quark momentum spectra, 
\ie, (i) pQCD distributions from PYTHIA, representing no rescattering, 
and, (ii) complete thermalization including collective expansion.   

\section{Coalescence into Charmonium}\label{coal}
Let us start by recalling the basic elements of the coalescence
model. In the Wigner formalism, the pertinent $p_T$-spectrum 
of a meson $M$ takes the form
\begin{eqnarray}
\frac{d^2N_M}{d{\bf p}_{\rm T}^2}= g_{M}\int \prod_{i=1}^{2}
\frac{p_{i}\cdot d\sigma _{i}d^{3} \mathbf{p}_{i}}{(2\pi )^{3} E_{i}}
 \ f_{q}(x_{1},p_{1}) \ f_{\bar q}(x_{2},p_{2}) \nonumber\\
\times f_{M}(x_1,p_1;x_2,p_2) \ 
\delta^{(2)}({\bf p}_{\rm T}-\bf{p}_{\rm 1T}-\bf{p}_{\rm 2T}) \ ,
\label{dndpt}
\end{eqnarray}
with $d\sigma_i$ denoting space-like hypersurface elements.
$f_{q,\bar q}(x_i,p_i)$ are invariant distribution 
functions of anti-/quarks depending on their space-time position 
and 4-momentum, and including spin and color degeneracy.  
The statistical factor $g_M$ accounts for the probability of forming 
a colorless meson of given spin from the underlying quark color and
spin (\eg, $g_{\psi}$=1/12), whereas $f_M$ encodes the dynamical
part of the process. In the following, we will adopt plane-wave 
single-quark distribution functions, and neglect
binding-energy corrections to $f_M$, in which case it becomes a Wigner
distribution function.  

\begin{figure}[th]
\includegraphics[height=2.8in,width=3.0in]{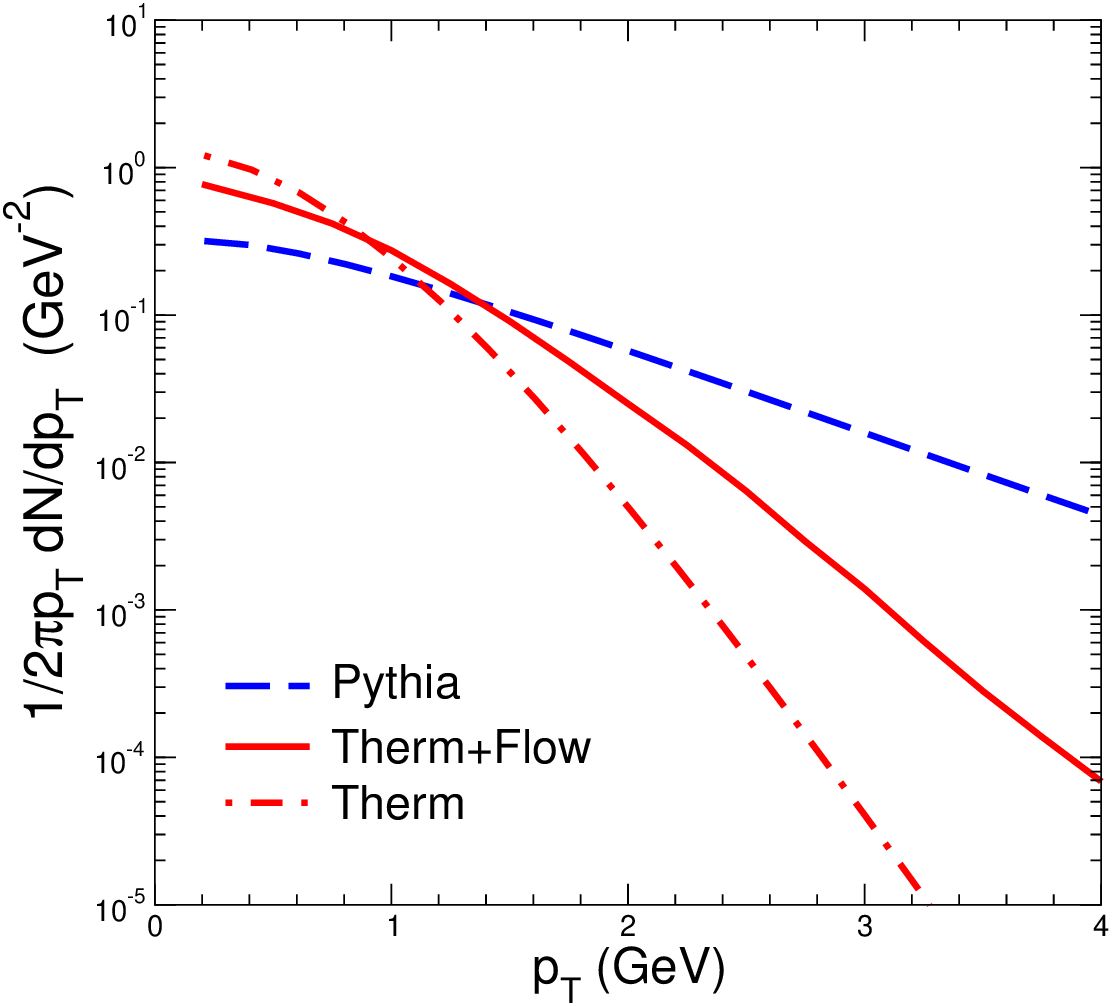}
\caption{
Transverse-momentum spectra of charm quarks in central 
200~AGeV Au+Au collisions at midrapidity for different scenarios:
static thermal (dash-dotted line), thermal plus transverse flow (solid line)
and primordial pQCD (dashed line).}
\label{fig1}
\end{figure}

Since the measured hadron distributions in heavy-ion collisions
at RHIC energies are approximately uniform around midrapidity, we 
will assume this also for the anti-/quark distribution functions, 
$f_{q,\bar q}$, with an additional Bjorken-type correlation between 
longitudinal coordinate and momentum~\cite{bjorken}. In the transverse 
direction, we follow Ref.~\cite{greco2} by employing light-quark
momentum distributions that are a combination of thermal spectra 
with radial flow and quenched minijets according to the GLV 
approach~\cite{glv} with an opacity parameter $L/\lambda=3.5$.  
For the transverse momentum distribution of charm quarks,
two possibilities will be investigated: 
(i) pQCD spectra as obtained from the PYTHIA event 
generator~\cite{pythia} and, (ii) thermal spectra at a temperature of 
$T$=170~MeV with radial flow. For the latter case, we take 
as an upper limit the same flow profile as determined in 
Ref.~\cite{greco} for light quarks, \ie, a linear increase with 
the position in the transverse plane,
\beq
\beta(r)=\beta_{\rm max}\frac{r}{R}
\eeq
with $R$=~8.1~fm, the radius of the 
firecylinder at hadronization, and $\beta_{\rm max}$=0.5 or 0.65, representing
ineherent uncertainties within our approach. 
Since heavy quarks are expected to 
suffer less energy loss in the QGP than light quarks\cite{DK01,djord}, 
we neglect charm-quark energy loss in the PYTHIA spectra in 
present study. In Fig.~\ref{fig1}, we show by the dashed line the 
charm-quark $p_T$-spectrum from the PYTHIA event generator and by the 
solid line that from the thermal scenario with collective flow.
For comparison, Fig.~\ref{fig1} also contains a thermal $c$-quark 
spectrum without collective flow (dash-dotted line).

The wave functions of the charmonium states are taken to be of Gaussian 
form in coordinate space, implying that the corresponding Wigner 
function is a Gaussian as well, \ie, 
\beq
f_M=8 \ \exp(- x^2/\sigma^2) \ \exp(- q^2 \sigma^2) \   
\eeq
with $x = x_1- x_2$, and $q^2=(m_1-m_2)^2-(p_1 -p_2)^2 $, where $x$ and 
$p$ are 4-coordinate and -momentum (see Ref.~\cite{greco2} for details). 
The width of the Gaussians is related to the mean-square-radius 
$r_{\rm rms}^2$=$<r^2>$ via $r_{\rm rms}^2=3\sigma^2/8$. Typical values 
from Cornell-type potential models\cite{eich} amount to 
$r_{\rm rms}$=0.47~fm for $J/\psi$, $0.74$~fm for $\chi_c$ and $0.96$~fm 
for $\psi^{\prime}$, which will constitute our baseline values. 

To illustrate basic features of the coalescence approach in a 
transparent way we first employ thermal $c$-quark spectra without 
radial flow. Under these conditions, the momentum spectrum, Eq.~(\ref{dndpt}),
can be evaluated analytically upon invoking the non-relativistic 
Boltzmann approximation for the $c$-quark distribution functions, 
suitable for charmonia at low and moderate $p_T$. One finds 
\beq
N_{\psi}\simeq g_{M }N_{c\bar c}^2\left[ \frac{%
(4\pi )^{\frac{3}{2}}\sigma ^{3}}{V_H (1+m_{c}T_H \sigma^{2})}\right]
\left(\frac{\tau^2}{\tau^2+m_c^2 \sigma^4}  \right)^{1/2} \ ,  
\label{npsi}
\eeq 
with $N_{c\bar c}$ the total number of $c\bar c$ pairs within 
the rapidity range $|y|\leq 0.5$. The bulk system is characterized 
by the proper time $\tau$, volume $V_c$  and temperature $T_c$ of the 
fireball at hadronization. Typical values for these quantities are 
$\tau$=4.3~fm, $T_c$=170~MeV, and $V_c$=900~fm$^3$, as determined in 
Ref.~\cite{greco} from numerical evaluations in the light-quark sector, 
which, after inclusion of transverse flow, reproduce well $p_T$-spectra 
of pions, protons, and kaons in central Au+Au 
collisions at 200~AGeV. The main features of Eq.~(\ref{npsi}) are:
(a) a weak dependence on the hadronization time as long as 
$\tau\ge$~2-3~fm/c and $r_{\rm rms}\le 0.5$~fm; (b) a ratio 
$\sigma^3/V_c \sim V_{\rm hadron}/V_c$ of hadron eigenvolume
over fireball volume, characteristic for coalescence (note, 
however, that the entailing increase of $J/\psi$ production with 
$\sigma$ is an artifact related to the neglection of binding energy 
effects; we will return to this issue below); (c) a dependence on
the number of participants, $A$, as 
$N_\psi/N_{c\bar c}\propto N_{c\bar c}/V_c\propto A^{4/3}/A\propto
A^{1/3}$, consistent with statistical 
models~\cite{pbm00,goren01,GR01,thews03}.

\section{Charmonium Spectra}
\label{charmonium}
The numerical evaluation of Eq.~(\ref{dndpt}) is performed along the 
lines of Ref.\cite{greco}. For each scenario, the total number of 
$c\bar c$ pairs is fixed to $N_{c\bar c}=2.5$ over one unit of 
rapidity. This number follows from an extrapolation of charm production
at fixed target energies to central Au+Au collisions at 200~AGeV,
corresponding to leading-order pQCD calculations upscaled by an 
empirical $K$-factor of $\sim$5~\cite{pbm98}, as well as the number of 
$N$-$N$ collisions. With a charm-quark mass of $m_c=1.4$~GeV,
the resulting open-charm cross sections are consistent with first 
indirect measurements via single electrons in 130 and 200~AGeV Au+Au 
by PHENIX~\cite{phenixe} at all centralities.

\begin{figure}[tbh]
\includegraphics[height=2.8in,width=3.0in]{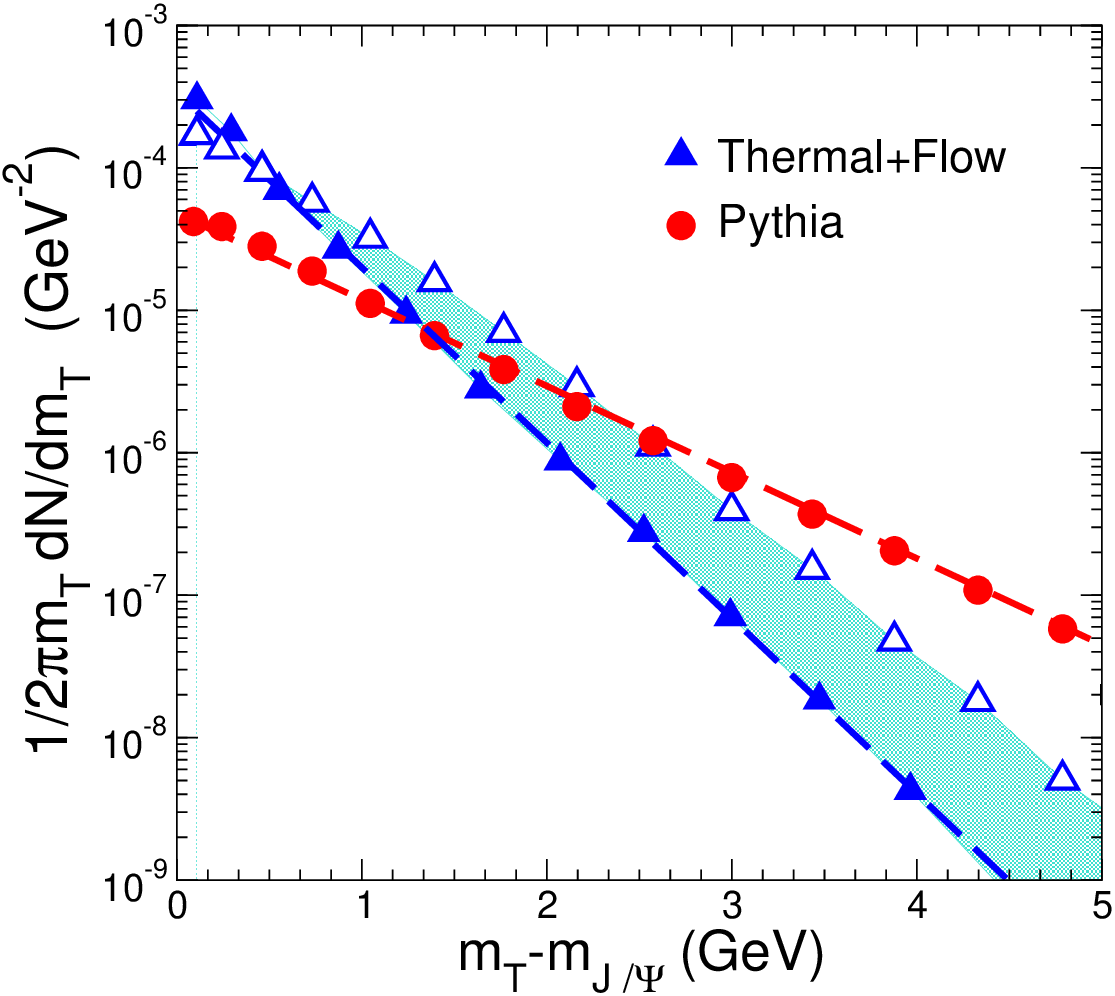}
\caption{$J/\psi$ transverse-mass spectra for central Au+Au
collisions at $\sqrt{s}=200$~AGeV from charm-quark coalescence:
circles correspond to PYTHIA $c$-quark distributions, while filled
and open triangles follow from thermal spectra with radial flow 
velocities of $\beta_{\rm max}$=0.5 and 0.65, respectively. Lines 
represent exponential fits where appropriate.}
\label{fig2}
\end{figure}
Fig.~\ref{fig2} displays our results for $J/\psi$ $m_T$-distributions
(without feeddown corrections); since, due to its large mass, 
the $J/\psi$ is more sensitive to variations in the collective
expansion then light hadrons, we indicated inherent uncertainties by 
using two values for the radial fireball-surface velocity at $T_c$, 
$\beta_{\rm max}$=0.5 and 0.65.   
We find that both pQCD-based (circles) and thermal $c$-quark spectra 
with $\beta_{\rm max}$=0.5 (filled triangles) exhibit 
approximately uniform slopes, which are quantified by the fitted lines 
to be $T_{\rm slope}$=720~MeV and $T_{\rm slope}$=350~MeV, respectively
(for $\beta_{\rm max}$=0.65, $T_{\rm slope}$=$550$~MeV in the region 
$m_T-m_{J/\Psi}=\,0-1.5$~GeV).
The average $<p_T>$ decreases from 1.95~GeV for the pQCD-based case
to a range of 1.29-1.60~GeV for the thermal one.  Furthermore, 
the total $N_{J/\psi}$ obtained 
by employing $c$ and $\bar c$ spectra from pQCD is smaller than that 
from the thermal model by about a factor of 3, as the flatter 
pQCD spectrum entails a substantial depletion of the probability 
in momentum space to form charmonia at low and moderate 
$p_T$ where their yield is concentrated. Contrary to the 
single-electron spectra from heavy-meson decays, which appear to be 
relatively insensitive to the difference between pQCD and 
thermal+flow open-charm distributions~\cite{Bats03} (at least up to 
$p_T\simeq$~3~GeV, and including ``contaminations" from $B$-meson 
decays), $J/\psi$ spectra exhibit a more pronounced sensitivity, which 
could enable it a rather direct window for studying the charm-quark 
behavior in the QGP. 

\begin{table}[t]
\caption{
Number of charmonia produced from charm-quark coalescence
at mid-rapidity in central collisions of Au+Au at $\sqrt{s}=200$ GeV.}
\medskip
\begin{tabular}{cccc}\hline\hline
\hfil & $\quad N_{J/\psi}\quad$ & $\quad N_{\chi_c}\quad$ &
$\quad N_{\psi\prime}\quad$ \\ \hline
Thermal & 2.6$\times 10^{-3}$ & 0.2$\times 10^{-3}$ & 0.7$\times 10^{-4}$ 
\\ \hline
PYTHIA & 0.8$\times 10^{-3}$ & 0.5$\times 10^{-4}$ & 0.2$\times 10^{-4}$ 
\\ \hline\hline
\end{tabular}
\label{tab:charm}
\end{table}
Let us briefly address the question of feeddown corrections from 
$\chi_c$ and $\psi^{\prime}$ resonances. A rigorous treatment of their 
formation requires the inclusion of in-medium effects on binding 
energies, charm-quark masses and radii in the coalescence probability, 
which is a much more involved problem. First studies
in this direction, including formation and dissociation processes, have 
been undertaken in Refs.~\cite{GRB03,thews03,zhang}.  For simplicity, 
we take as a guideline thermal mass weights according to   
$(m_R/m_{J/\psi})^{3/2}\exp(-(m_R-m_{J/\psi})/T)$ for their
production relative to that for $J/\psi$.
The $J/\psi$ numbers corresponding to the spectra in Fig.~\ref{fig2} are 
collected in Table~\ref{tab:charm}, together with estimated feeddown 
corrections. Despite the significantly larger radii of $\chi_c$ and 
$\psi^{\prime}$ states, their combined contribution to $J/\psi$ is 
still only $\sim$10\%. A more precise assessment is beyond the scope 
of this work. However, we expect the relative suppression of the $J/\psi$ 
yield and its harder slope, as obtained with the pQCD charm-quark 
spectrum  compared to the thermal+flow case, to 
be a rather robust result. It could prove valuable in,  
\eg, disentangling a transition from coalescence of soft $c$-$\bar c$ 
to hard ones (or even unsuppressed primordial charmonia) in $p_T$,
as well as in probing the degree of thermalization of charm quarks in 
the QGP.

\section{$D$-Meson Spectra}
\label{charm}
The standard coalescence approach becomes unreliable if the phase space
density of either one (or both) of the quarks is no longer small, such 
as in bulk hadron production at low $p_T$. For the case of interest 
here, \ie, $D$-meson spectra, we expect the following results to be 
trustworthy for $p_T\ge$~1~GeV. 
Again, our calculations proceed along the lines of Ref.~\cite{greco}, 
with the same fireball parameters as used above, \ie, a constituent 
light-quark mass of 300~MeV, and the two scenarios for charm quarks 
(pQCD vs. thermal+flow). The width of the Gaussian Wigner function 
for $D$'s is taken to render their radius to 0.6~fm.  
We have again weighted the production of $D^*$'s with 
a thermal factor $(m_{D^*}/m_D)^{3/2}\exp(-(m_{D^*}-m_D)/T)$
relative to that of $D$ mesons. Semileptonic decays of $D$-mesons
have been computed assuming predominance of 3-body decays, 
$K\nu e$ and $K^*\nu e$, with a phase-space weighting according to the
weak matrix element. 

\begin{figure}[th]
\includegraphics[height=2.8in,width=3.0in]{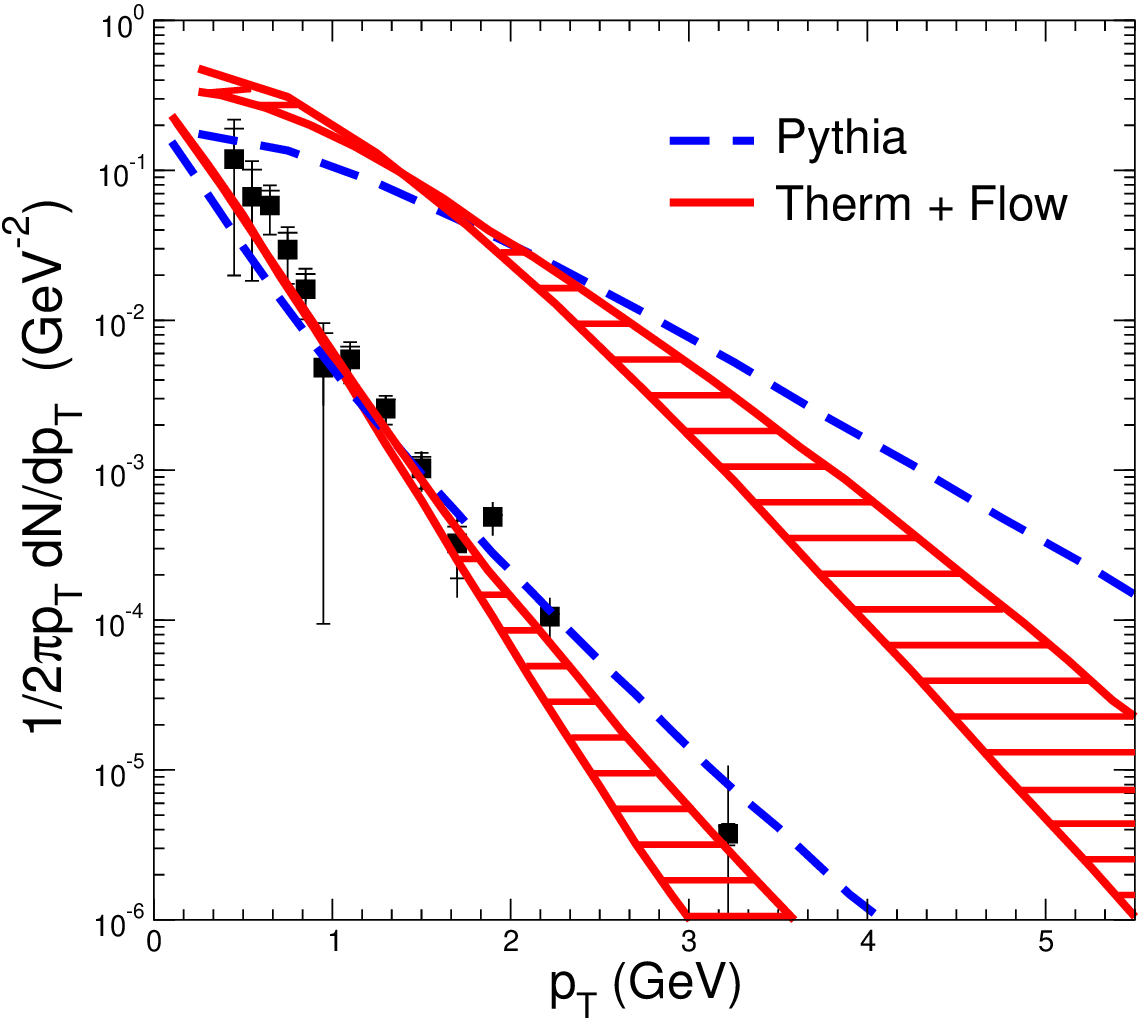}
\caption{Coalescence results for transverse-momentum spectra of 
all $D$-mesons (upper curves) and pertinent electron decays (lower 
curves) in central 200~AGeV Au+Au collisions at midrapidity.
Dashed lines are obtained from pQCD $c$-quark distributions, while 
shaded bands correspond to thermal spectra with flow velocities
between 0.5 and 0.65. ``Non-photonic" single-$e^\pm$ data 
(squares with both statistical and systematic errors)
are from PHENIX~\cite{phenixe}.}
\label{fig3}
\end{figure}

In Fig.\ref{fig3}, the upper curves display the coalescence 
results for the $p_T$-spectra of charmed mesons $D$, $D^*$, $D_s$ and 
$D_s^*$. Also for $D$-mesons, the pQCD $c$-quark distributions induce 
harder spectra (dashed line) than the 
thermal+flow ones, although somewhat less pronounced than for the 
charmonium case; the shaded bands in Fig.\ref{fig3} cover the range of 
flow velocities from $\beta_{\rm max}$=0.5 to $0.65$).  
The lower curves show single-$e^\pm$ spectra from decays of all
$D$- and $D_s$-mesons (including feeddown from $D^*$'s). Similar 
to Ref.~\cite{Bats03}, we find that, on the basis of the current 
experimental accuracy, also within a coalescence framework, underlying 
pQCD $c$-quark spectra cannot really be distinguished from thermalized 
ones with flow. Our coalescence spectrum from thermal
$c$-quarks with flow indeed closely resembles the hydrodynamic 
$D$-meson spectra of Ref.~\cite{Bats03} (implying that, modulo charmed baryons,
essentially all $c$-quarks recombine). This is also true when comparing
the pQCD-based $c$-quark coalesence with the fragmentation spectrum in 
Ref.~\cite{Bats03} above $p_T$$\simeq$2~GeV. 
Below $p_T\simeq$~2~GeV, however, there is a significant lack in the 
yield of the pQCD coalescence spectrum as compared to  
inclusive PYTHIA $D$-mesons, since no fragmentation contribution has been 
accounted for in the present study.

Concerning the coalescence yields for $D_s$ mesons
(including feeddown from $D_s^*$), we find for the two cases of 
thermal and pQCD $c$-quark spectra $D_s$/$D$ ratios of 29\% 
and 25\%, respectively, with spectral shapes very similar to 
the inclusive ones of Fig.~\ref{fig3}. These values are quite 
reminiscent to the 23\% found within the coalescence approach of 
Ref.~\cite{RS03} for $p$-$p$ collisions at RHIC.

\section{charm and single-electron elliptic flow}\label{elliptic}
Finally, we turn to the elliptic flow, $v_2$, of charmed particles,
which is a well-established probe of the degree of thermalization
in the (early) QGP as it vanishes in the absence of any rescattering. 
The coalescence model has been shown to quantitatively 
describe~\cite{greco2,fries2,linv2} the elliptic flow of light
hadrons such as pions, protons, kaons and lambdas. 
Therefore, we believe that its pertinent predictions especially for 
$D$-mesons, and to a lesser extend for charmonia, are a rather  
reliable measure of charm-quark reinteractions.
Even if $c$-quarks do not experience any final-state interactions, 
a lower limit for the $v_2$ of $D$-mesons from coalescence arises 
due to the $v_2$ of their light-quark component. On the contrary, an 
upper limit can be expected from the assumption that the charm quark 
has the same $v_2^{\rm max}$ as the light quarks.

Some specific features of charmed hadron $v_2(p_T)$ have already been 
pointed out within a schematic calculation in Ref.~\cite{lin03}.
In particular, it has been shown that even if $c$ quarks acquire
the same $v_2(p_T)$ as light quarks, the large difference in the 
quark masses entails a smaller value of $v_2$ for
$D$-mesons than for $J/\psi$ at same $p_T$.  
This is so because the $D$-meson momentum is largely determined 
by the $c$-quark, and an equal velocity of $c$ and light quark,
as required for coalescence, implies the light quark to be at 
low momentum where its $v_2$ is relatively small. 
Here, we quantify this study by including more realistic quark
distribution functions, especially through the inclusion
of radial flow, which by itself affects light and 
heavy quarks in a different way, and extend it to the semileptonic
decay electrons.  

\begin{figure}[tbh]
\includegraphics[height=3.0in,width=3.0in,angle=0]{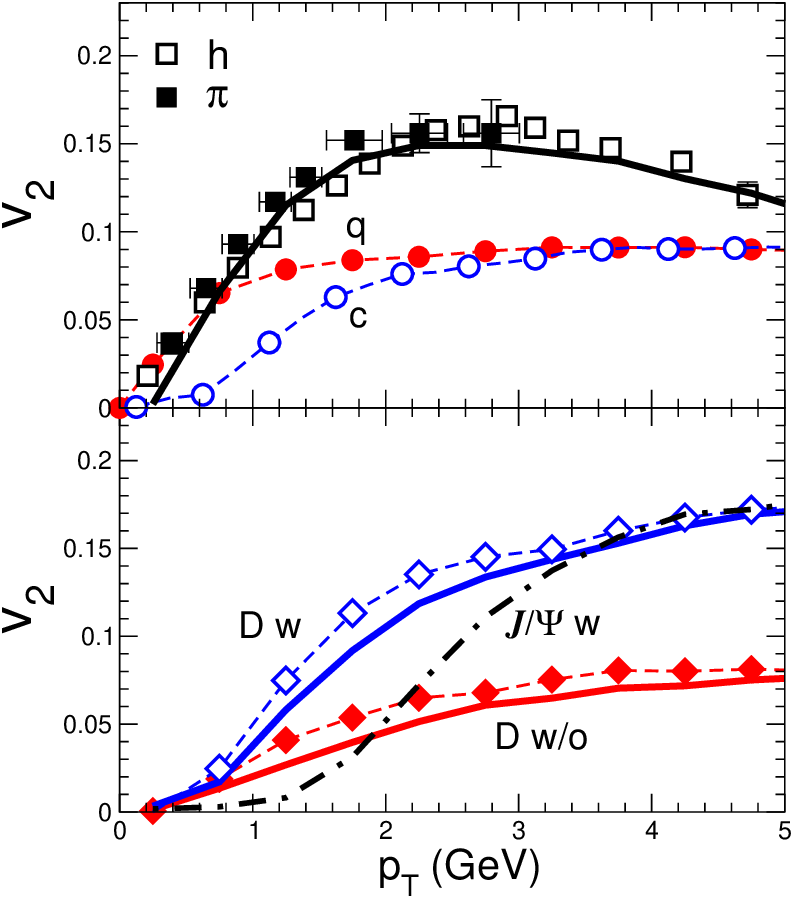}
\caption{Elliptic flow in minimum bias Au+Au collisions
at $\sqrt{s}$=~200~AGeV. Upper panel: input parameterizations for 
light (solid circles) and charm quarks (open circles) together with 
the $v_2$ for pions (solid line) from coalescence plus fragmentation, 
as well as experimental data from PHENIX for $\pi^{+}$ (filled 
squares)\cite{esumi} and charged particles (open squares)~\cite{adlv2}.
Lower panel: $D$-mesons (solid lines, lower: using pQCD charm-quark 
distributions, upper: for thermal+flow $c$-quark spectra at $T_c$)
and respective decay electrons (diamonds), 
as well as $J/\psi$ mesons (dash-dotted line).}
\label{fig4}
\end{figure}

Fig.~\ref{fig4} summarizes our elliptic flow results. For the upper 
panel, we recall the underlying light-quark $v_2$ (solid circles), 
which has been fit (solid line) to reproduce the experimental data for pions 
and charged hadrons (filled and open squares, respectively) 
in Ref.~\cite{greco}, together with a charm-quark $v_2$ (open circles)
saturating at the same maximal value (and with equal radial flow).
The lower panel contains the ensuing results for $D$-mesons (solid 
lines) and their decay electrons (diamonds), as well as $J/\psi$'s
(dash-dotted line). For $D$-mesons, the elliptic flow from 
the two scenarios of complete thermalization and no reinteractions 
of the $c$-quarks deviates by about a factor of 2 for $p_T$$\ge$1.5~GeV. 
The decay $e^\pm$ are found to essentially preserve the $v_2$ of 
their parent particles, which makes it a highly interesting observable 
for upcoming measurements.  We note that $B$-meson contributions
have not been included in present study. If these become significant
for $e^\pm$ momenta above 2~GeV, caution needs to be exercised 
in drawing conclusions about $c$-quark elliptic flow. 

For the elliptic flow of $J/\psi$'s, which entirely stems from 
(thermalized) $c$-quarks, we observe a significant delay of
its increase with $p_T$ as compared to that for $D$-mesons, 
caused by the radial flow. Indeed, if the latter is absent, the $v_2$
of $J/\psi$'s saturates faster than for $D$-mesons, see, 
\eg, Ref.~\cite{lin03}.
We furthermore recall that, on the one hand, the $v_2$ of $J/\psi$'s is 
zero for coalescence between pQCD charm quarks. On the other hand, we 
have not taken into account any primordially produced $J/\psi$'s, for 
which suppression mechanisms, if active early in the evolution, can 
induce a nonzero $v_2$~\cite{WY02}. However, the magnitude of such 
effects (which can arise from both nuclear absorption and QGP dissociation)
has been estimated to be rather small, up to $\sim$2\% (compared to 10\% or 
more for coalescence between thermalized $c$-quarks).  
Nevertheless, in a more complete description, one expects an increasing
fraction of (unsuppressed) primordial charmonia towards higher
$p_T$, thus reducing $v_2$.

\section{summary}\label{summary}
Within a coalescence approach as successfully applied earlier in 
the light-quark sector, we have evaluated transverse-momentum 
dependencies of charmed hadrons in central heavy-ion reactions at 
RHIC.  For the charm-quark distributions at hadronization
we have considered two limiting scenarios, \ie, no 
reinteractions (using spectra from PYTHIA) and complete 
thermalization with transverse flow of the bulk matter.  
The resulting $J/\psi$ ($m_T$-) spectra differ in slope by up to a 
factor of 2 (harder for pQCD $c$-quarks), and the integrated yield 
is about a factor of 3 larger in the thermal case. For $D$-mesons, we 
found that the difference in the slope parameters of the $p_T$-spectra 
in the two scenarios is less pronounced, but their elliptic flow 
is about a factor of 2 larger for $p_T$$\ge$1.5~GeV in the thermalized 
case. The elliptic flow pattern of  $D$-mesons was found to be essentially 
preserved in the single-electron decay spectra, rendering the latter a very 
promising observable to address the strength of charm reinteractions 
in the QGP. The present study can be straightforwardly generalized to 
charmed baryons ($\Lambda_c$), which may serve as a complimentary probe 
for charm-quark reinteractions in the QGP.

\section{acknowledgments}
This work was supported in part by the US National
Science Foundation under Grant No. PHY-0098805 and the Welch
Foundation under Grant No. A-1358. VG was also supported by 
the National Institute of Nuclear Physics (INFN) in Italy.

\end{document}